\documentclass[a4paper]{article}
\usepackage{interspeech-2018/INTERSPEECH2018,amsmath,graphicx}
\usepackage{subcaption, dsfont}

\title{Deep Speech Denoising with Vector Space Projections}
\name{Jeffrey Hetherly, Paul Gamble, Maria Barrios, Cory Stephenson, Karl Ni}
\address{Lab41, In-Q-Tel}
\email{\{jhetherly, pgamble, mbarrios, cstephenson, kni\}@iqt.org}
\begin{document}
\ninept
\maketitle
\begin{abstract}
We propose an algorithm to denoise speakers from a single microphone in the presence of non-stationary and dynamic noise. Our approach is inspired by the recent success of neural network models separating speakers from other speakers and singers from instrumental accompaniment. Unlike prior art, we leverage embedding spaces produced with source-contrastive estimation, a technique derived from negative sampling techniques in natural language processing, while simultaneously obtaining a continuous inference mask. Our embedding space directly optimizes for the discrimination of speaker and noise by jointly modeling their characteristics. This space is generalizable in that it is not speaker or noise specific and is capable of denoising speech even if the model has not seen the speaker in the training set. Parameters are trained with dual objectives: one that promotes a selective bandpass filter that eliminates noise at time-frequency positions that exceed signal power, and another that proportionally splits time-frequency content between signal and noise. We compare to state of the art algorithms as well as traditional sparse non-negative matrix factorization solutions. The resulting algorithm avoids severe computational burden by providing a more intuitive and easily optimized approach, while achieving competitive accuracy.
\end{abstract}

\noindent\textbf{Index Terms}: deep learning, speech, speaker denoising, non-stationary processes

\section{Introduction}
\label{sec:intro}

Signal denoising has been a problem in multiple media for over a century with applications ranging from acoustic speech processing, image processing, seismic data analysis, and other modalities. For each application, approaches have evolved over the span of several decades ranging from traditional statistical signal processing like Wiener and Kalman filtering, wavelet theory, and specific instances of matrix factorization. While effective for locally as well as wide-sense stationary signals, even with a storied history, these efforts have seen less success with more dynamic and \emph{in-the-wild} sets of noise owing to their algorithmic capacity. 

Dynamic noise represents many real-world speech situations, and solutions that have impressive results primarily focus on hardware: array processing efforts~\cite{bss-beamforming} in the form of SONAR, RADAR, and Synthetic Aperture sensing. These methods solve the problem by using the inputs of multiple sensors to process the source of interest. Unfortunately, much of recorded contemporary media is typically done through a phone, and the same methods cannot be easily extended to the monaural case~\cite{monaural-ica}, in which audio is recorded from a single microphone.

In previous approaches to the monaural problem, assumptions are explicitly made on signal and noise attributes, or specifications have enabled some control over environment and listening devices.
The more general case of single-track speech recordings in noisy or reverberant rooms has become increasingly common due to the proliferation of inexpensive portable devices with recording capabilities such as cellphones and laptops. In such cases, no guarantees on speech or noise attributions can be assumed regarding the nature of the environment or the location of microphones. 

Over the last decade, machine learning approaches have begun to see success in this scenario.
In particular, adaptation of familiar matrix factorization techniques~\cite{sparse-nmf} to the processing of time-frequency representations of audio signals has proven useful.
However, these methods can be difficult to make performant~\cite{le2015sparse}, and in many cases additional complexity is required to model source characteristics accurately.

More complex sources can be modeled with the inclusion of \emph{a priori} knowledge regarding their characteristics.
This can be empirically derived from a large corpus of training data provided the model in use has a high capacity.
In recent years, neural network approaches have reemerged in the mainstream of machine learning research, and several papers~\cite{wavenet} have adapted them to the general speech denoising problem. Among these methods, recurrent neural networks in particular have shown the most promise in modeling acoustic time series~\cite{hershey2016deep, 2016chimera}, especially when applied to time-dependent spectral features.

One challenging aspect of neural network approaches is the development of cost functions. For speech signals in particular, the computational complexity of the cost function is important as the timescales associated with speech contain many samples.
Additionally, if the goal is to separate sources (e.g, a speaker and a noise source), the cost function must be invariant to different permutations of the recovered sources since the ordering is arbitrary.
The proposed approach automates the featurization of speakers and the characterization of noise using an efficient permutation invariant sampling technique.


Building upon previous work, we propose an algorithm to \emph{directly} optimize a vector space that isolates specific source characteristics.
Such an approach is closely related natural language processing~\cite{mikolov} work, where an embedding space in which speakers and noise sources are explicitly contrasted with each other is created.
The conjecture is that such an intuitive approach will provide better discrimination, and the proposed algorithm source-contrastive estimation (SCE), as distinguished from noise-contrastive estimation~\cite{gutmann2010noise}.
Our vector space is independent of source type and offers a significant speedup at training time compared to state of the art deep-clustering-based approaches. Additionally, we further improve upon our model with a mask inference approach detailed in~\cite{2016chimera}.

The remainder of this paper describes our approach to finding optimal vector spaces using SCE and applies the technique to synthetically-mixed noisy speech.
We dive deeper into the state of the art techniques, some of which we leverage, in Sec.~\ref{related}.
The approach is then described in Sec.~\ref{approach} with implementation details in Sec.~\ref{implementation}.
Experimental results are shown in Sec.~\ref{results}, which is followed by a summary and discussion of future work.
\section{Related Work}
\label{related}

Considerable work has been done in speech processing for as long as recordings have been made. Traditional methods have included approaches that are rooted in signal processing theory, where a large number of approaches use some type of matrix factorization
~\cite{Kameoka2016, monaural-nmf}.
In particular, sparse non-negative matrix factorization (SNMF) is shown effective at extracting non-stationary noise sources in~\cite{le2015sparse, schmidt-nmf, sparse-nmf}.
SNMF constructs a set of spectral basis functions from training data and linearly combines these with a set of learned weights to reconstruct the spectral features of the desired signal.
Sparsity is typically enforced by an $\ell_1$-norm constraint on the learned weights that contains a multiplicative hyper-parameter, $\mu$.
As will be shown in Sec~\ref{results}, linear methods such as these lack the algorithmic capacity to compete with more modern techniques.

\subsection{Convolutional Denoising Autoencoder}

Autoencoders have been used to successfully remove noise and to isolate single sources from audio signals~\cite{feng2014speech}.
At a high level, autoencoders learn to featurize inputs (usually referred to as encoding) and then reconstruct them as outputs (decoding). 
This approach is well suited to denoising because a model is forced to preferentially build representations of the non-noise components of its input.

Closely related are convolutional autoencoders, typically used for denoising images~\cite{2017arXiv170308019G, vincent2010stacked}. These models make use of convolutional layers during encoding and deconvolutional layers during decoding. Convolutional denoising autoencoders (DAE) applied to audio signals represented via spectrogram (using STFT) operate similarly, though many of these approaches have problems generalizing to unseen signals. Moreover, convolutional autoencoders are an architectural construct still dependent on their cost functions, a challenge that defines how they perform in the context of denoising, where the common $\ell_2$-norm may not be sufficiently descriptive. 



\subsection{Neural Network-Based Source Embeddings}

Recent success in monaural audio source separation and denoising have taken to learned embedding vectors~\cite{hershey2016deep, isik16, attractor, 2016chimera}.
The primary advantage of learning embedding vectors is that they bypass the so-called permutation problem in which the output of a learning algorithm must be permuted to account for the unordered nature of the target sources~\cite{dong-permutation}.
Additionally, the number of sources to be separated and denoised can be arbitrary with an appropriate clustering technique (although, this depends on how inference is performed).

The embedding model we propose in the following sections most resembles deep clustering~\cite{hershey2016deep} and mask inference (DC+MI) found in~\cite{2016chimera}, though with a vastly reduced cost function. 
The DC+MI network learns embeddings given the spectral magnitude of the mixed audio sample using a series of four bi-directional LSTMs (BLSTM).
In addition to clustering those embeddings to create a binary mask as in~\cite{hershey2016deep}, a learned non-linear transformation is used to directly translate the embeddings into a ratio mask. This has the advantage of limiting some of the artifacts inherent to performing a binary mask.
However, this comes at a cost of fixing the number of sources to two. Clustering on an arbitrary number of sources can still be performed on the embeddings, but only a binary mask can be constructed from these clusters.

\section{Approach}
\label{approach}
The approach used in this paper combines the mask inference capabilities of cited literature in Sec.~\ref{related} with the flexibility of SCE to remove dynamic, non-stationary noise sources from speech for monaural audio signals.

\subsection{Datasets}

Our task is to isolate speech from a mixture of dynamic noise and speech using a monaural audio signal.
All denoising algorithms are trained and evaluated on a mixture of the LibriSpeech~\cite{librispeech} and UrbanSound8K~\cite{UrbanSound2014} datasets.
LibriSpeech provides high-quality audio recordings of isolated English speech from both male and female speakers and UrbanSound8K provides recordings from ten non-stationary noise classes.
Two two-second clips from each dataset are added at various SNR ratios to create the noisy-speech data.
The SNR ratio is continuously varied between -5 and 5 dB for the training phase for all but SNMF algorithm wherein speech and noise are fed in separately.
No impulse response convolution is used so as to focus solely on removing non-stationary sources of noise.

For the training, validating, and in-set testing of each algorithm we use the \texttt{train-clean-100} set of audio readings from the LibriSpeech dataset, which provides approximately 100 hours of speech evenly split between female and male speakers.
For out-of-set testing we use the \texttt{dev-clean} set from LibriSpeech.
Although all noise types from UrbanSound8K are used for training, noise files from each noise type are reserved for training, validation, and testing.

\subsection{Model}

Our model for denoising monaural signals operates on the assumption that linearly mixed speech and noise $x(t)$ can be well-separated into individual sources $s_i(t)$.
In this context, a source is either a speaker or a particular type of noise. For a given source $i$ in a speaker-noise mix, our model masks the magnitude response.
This mask filters out information from time-frequency bins in the short-time Fourier transform (STFT), $X(t,f)$, that do not belong to a given source, while passing those time-frequency bins that do. 

Typically, the predicted mask $Y_{t,f}^{(i)}$ for the $i^{th}$ source is implemented as either a ratio or in our case, a binary mask.
We let $Y_i \in \{-1, 1\}^{M \times TF}$, where $M \leq C$, $C$ being the total number of sources in our training set and $M$ being the number to be mixed.
To set our masks, if $i^{th}$ source is the loudest in time frequency bin $(t,f)$, then $Y_{t,f}^{(i)}=1$, and $Y_{t,f}^{(i)} = -1$ otherwise.

\begin{figure}[h]
\centering
\includegraphics[width=0.45\textwidth]{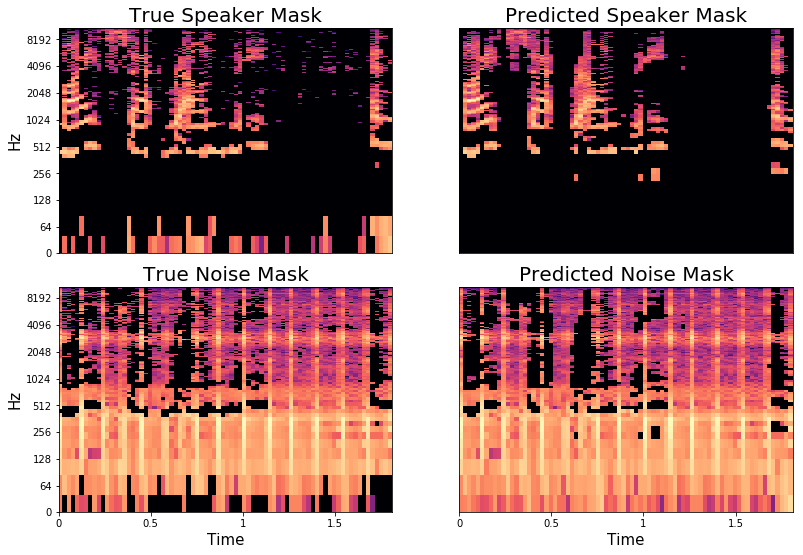}
\caption{Example of predicted and true binary masks applied to a sample mixed at an SNR of -1}
\label{algorithm}
\end{figure}


Similar to natural language processing embedding techniques like \emph{word2vec}~\cite{mikolov}, a given word embedding can represent specific words.
Instead of a word embedding, we use a speaker embedding, similarly optimized via two vector spaces.
The first vector space is an input embedding that \emph{implicitly} defines a speaker, and it is not associated with anyone in particular.
We also have an output embedding that \emph{explicitly} trains to a corpus of known speakers.
Then, when performing inference to input vector space, it is possible to generalize to any possible speaker by clustering our neural network outputs.
In our notation, the input and output vector spaces for a given sample are implemented as tensors with an embedding space of $E$, labeled as $V_i(t,f)$ and $V_o$, respectively.
The columns of either tensor have $E$ dimensions (hidden units) and denote the vectors associated with a speaker's likeness.

To train and generate our embeddings, we use a recurrent neural network regression to $V_i$.
To compare to~\cite{hershey2016deep} and~\cite{2016chimera}, we use a total of four BLSTM layers, and we have a dense layer that is convolved over the output 2D vector produced by the final BLSTM.
This final layer of source embeddings is also fed through a non-linear transform as in~\cite{2016chimera} to yield the ratio mask. 

Let our loss for every time frequency bin for sample $b$ be denoted as $\mathcal{L}^{(b)}_{t,f}$. Then,
\begin{equation}
\label{eqn:cost}
\mathcal{L}^{(b)}_{t,f}(\mathbf{v}_i, \mathbf{v}_o) = \frac{-1}{M} \sum_{s \in S_b} \log \sigma \left( Y^{(s,b)}_{t,f} \cdot \textbf{v}^{(b)}_i(t,f)^T \textbf{v}^{(s)}_o \right)
\end{equation}

Here, $S_b$ is the set of sources sampled for mix $b$, and $s$ is a single source from the subset.
The total loss for the batch of size $B$ over all frequencies and time is thus,

\begin{equation}
    \mathcal{L}(\mathbf{v}_i, \mathbf{v}_o) = \frac{1}{B} \sum_b \sum_{(t,f)} \mathcal{L}^{(b)}_{t,f}
\end{equation}

Intuitively, the output of the neural network at time $(t,f)$ is $\textbf{v}_i(t,f)$ and the output vector $\textbf{v}^{(s)}_o$ is an embedding for source $s$ at frequency $f$.
Say that source $1$ is louder than source $2$ at time frequency bin $(t,f)$ for sample $b$.
Then we would ideally like the correlation between the embedding produced by our neural network $\textbf{v}_i$ and the vector for source 1 to be high.
That is to say, we would like $\sigma (\textbf{v}_i^T \textbf{v}^{(1)} ) \rightarrow 1$.
Simultaneously, the correlation between $\textbf{v}_i$ and the vector for source 2 should be low, since these two vectors should be anti-correlated if they are sufficiently different.
That is to say, we would like $\sigma(\textbf{v}_i^T \textbf{v}^{(2)}) \rightarrow 0$.
Mathematically speaking, we are pulling our embedding towards our source vector $\textbf{v}^{(1)}_o$ and pushing it away from non-source vectors $\textbf{v}^{(2)}_o$.
Which sources to attribute appropriate correlation/anti-correlation to is determined by the label $Y$, which will be $+1$ in the former case and $-1$ in the latter.
It is important to note that we can save on both computation and accuracy by optimizing only those sources that are in $S_b$, which in our case will have two elements (one speaker and one noise source).


Additionally, during inference we do not use the output vector space $V_o$.
While it is true that computations are further reduced, the intention is that the out-of-set sources set is allowed.
In fact, even though we may train on mixes with fewer sources, we can inference in situations where there are arbitrary numbers of sources.


Our algorithm (denoted SCE+MI) is implemented in Tensorflow, v1.4~\cite{tensorflow}, with an architecture consisting of four BLSTM layers $r_1, r_2, r_3, r_4$ of 500 units each.
These are followed by a fully connected layer $d_1$ that maps the output of the fourth BLSTM layer to the input vector space.
The BLSTM layers use tanh nonlinearities, and the fully connected layer is linear.
For a batch of inputs $\mathbf{X}$, the output of the four BLSTM layers $r_1, r_2, r_3, r_4 \in \mathds{R}^{B \times T \times 500}$.
While the final (embedding) layer of the neural network is technically a fully-connected linear layer, it is implemented as a $1D$ convolution over the $r_4$ output tensor with a filter $w\in\mathds{R}^{1\times500\times F\cdot E}$.
The output of the convolution can then be reshaped to give the input vector space $\mathbf{V}_i\in\mathds{R}^{B\times T \times F \times E}$.
The vector-space output is fed through what is effectively a 1D convolution along the embedding dimension with a softmax that yields the final ratio mask output.
This implementation allows the model to be run for arbitrary input $T$, which is useful at inference time.

For efficient evaluation of the cost function of Eq. \ref{eqn:cost} across batches, the sources vectors for sources only represented in each batch are assembled into a tensor $\mathbf{V}_o\in\mathds{R}^{B \times M \times E}$.
The ordering of the $M$ speakers in $\mathbf{V}_o$ must match the ordering used in $\mathbf{Y}$, but is otherwise arbitrary.
To efficiently compute the dot products $V_i\cdot V_o$ in Eq. \ref{eqn:cost} with broadcasting, we expand the dimensions appropriately.


This gives an output of the dot product operation as a tensor $\mathbf{D}\in\mathds{R}^{B\times T\times F \times M}$, which is compatible with the labels $\mathbf{Y}$ and so they can be multiplied together elementwise to give the argument of the sigmoid in Eq. \ref{eqn:cost}.
The remaining portion of the cost function is easily evaluated.

Our batch size is $B=256$ during training.
The input tensors have dimensions $\mathbf{X}\in\mathbb{R}^{B \times T \times F}$ and label tensors are $\mathbf{Y}\in\mathbb{R}^{B \times T \times F \times M}$, where $T=78$ is the length of total time steps per sample and $F=257$ are the number of frequency bins used.
%
\section{Experiments}
\label{implementation}
%
In all experiments, signals are resampled and scaled to $10kHz$, zero mean and unit standard deviation, from which the 
short time Fourier transform (STFT) spectrograms are extracted with a Hanning window of 512 and 256 stride length.
We use audio clips of approximately two seconds which, when combined with the STFT operation, yield input features of dimension $257\times78$ (frequencies by time frames).
The complex phases $\phi_{t,f}$ were saved separately for use in post separation processing.
Separate spectrograms for the signal from each speaker and noise ($S^{(n)}_{t,f}$ for $n\in\{1,2,...,C\}$) were computed for training and evaluation purposes, while the total spectrogram was computed by the elementwise sum $X_{t,f}=S^{(n)}_{t,f}+S^{(m)}_{t,f}$ for a speaker and noise with IDs $n,m$.

The magnitude of the $X_{t,f}$ spectrograms were then passed through a square root nonlinearity and percent normalized.
This is similar to the procedure suggested in \cite{wang2014training}; however we obtained better results with a square root rather than a logarithmic nonlinearity.
%
Source labels $Y_{t,f}^{(c)}$ are assigned to each T-F bin by giving a value of $1$ to the signal which is loudest in at that time and frequency, and a value of $-1$ to all other sources.  

\subsection{Algorithm Comparisons}

We compare three approaches against the proposed work: a linear matrix factorization method (SNMF), a denoising auto-encoder (DAE), and a hybrid deep clustering/mask inference architecture (DC+MI).
SNMF is adopted from~\cite{le2015sparse} with the most optimal hyperparameter settings found therein and trained on 10000 two-second audio clips of noise and speakers.
To aid in training SNMF we removed portions of each spectrogram based on a log max-amplitude threshold for each time frame.
This threshold was found to isolate spoken words while trimming the surrounding empty audio.

Our comparison convolutional DAE is based on ~\cite{2016arXiv160608921M} and consists of 15 convolutional layers, followed by 15 deconvolutional layers.
Each layer contains 128 5x5 filters with relu activation and constant input size.
Skip connections are employed between every other pair of matched convolutional and deconvolutional layers.
The model was trained using Adam RMSprop with Nesterov momentum and a learning rate of 5e-5. 

The DC+MI network is an implementation of the architecture found in ~\cite{2016chimera}.
We use the same optimal set of hyperparameters as they do except that our loss function for the mask inference head uses the true spectral component rather than a proxy.

Comparisons of the performance of each alogrithm are quantified by improvement in the source-to-distortion ratio (SDR).
Each algorithm is evaluated on how well it improves the SDR metric for an input SNR range of $[-5,5] \mathrm{dB}$ and for each noise type. 

\subsection{Reconstruction}

At inference time for our model and the deep-clustering head of DC+MI, a signal consisting of an unknown mixture of sources is preprocessed as described in the previous subsection, giving a complex T-F estimate of a single source signal, $\hat{S}_{t,f}$.
An input feature is generated and fed through the model to obtain the vectors $\mathbf{V}_i$.
A $K$-means clustering is then performed on the vectors in order to generate a labeling prediction $\mathbf{\hat{Y}}\in\mathds{R}^{T\times F\times K}$ in which each T-F element is associated with a cluster label.
Here the element $\hat{Y}^{(k)}_{t,f}=1$ if the associated vector $V_{t,f}$ belongs to the $k^{th}$ cluster, and $\hat{Y}^{(k)}_{t,f}=-1$.
These labelings can then be used as masks to reconstruct a source $S^{(k)}_{t,f}$ from each of the $K$ clusters.
T-F representations of the inferred sources are calculated as the element-wise multiplication of the input spectrogram with the inferred labeling.
\begin{equation}
\hat{S}^{(k)}_{t,f} = X_{t,f}\odot \frac{1}{2} \left( \hat{Y}^{(k)}_{t,f} + 1 \right)
\end{equation}
The source spectrogram $\hat{S}^{(k)}_{t,f}$ is then converted (using the inverse STFT) into a source waveform, completing the inference process.

The output of the mask inference head of SCE+MI and DC+MI and that of SNMF is a ratio mask that, when multiplied element-wise with the original spectrogram, yield the respective speaker and noise sources.
These ratio masking techniques have the potential to produce higher-quality audio than binary masking as the T-F bins can be shared amongst sources (as is actually the case).


Our contribution, replicated research~\cite{2016chimera,le2015sparse}, and evaluation code can be found 
{\small \verb"http://github.com/lab41/magnolia"}. 
\section{Results}
\label{results}
\begin{figure}[!ht]
	\centering
	\begin{subfigure}{0.4\textwidth} 
		\includegraphics[width=\textwidth]{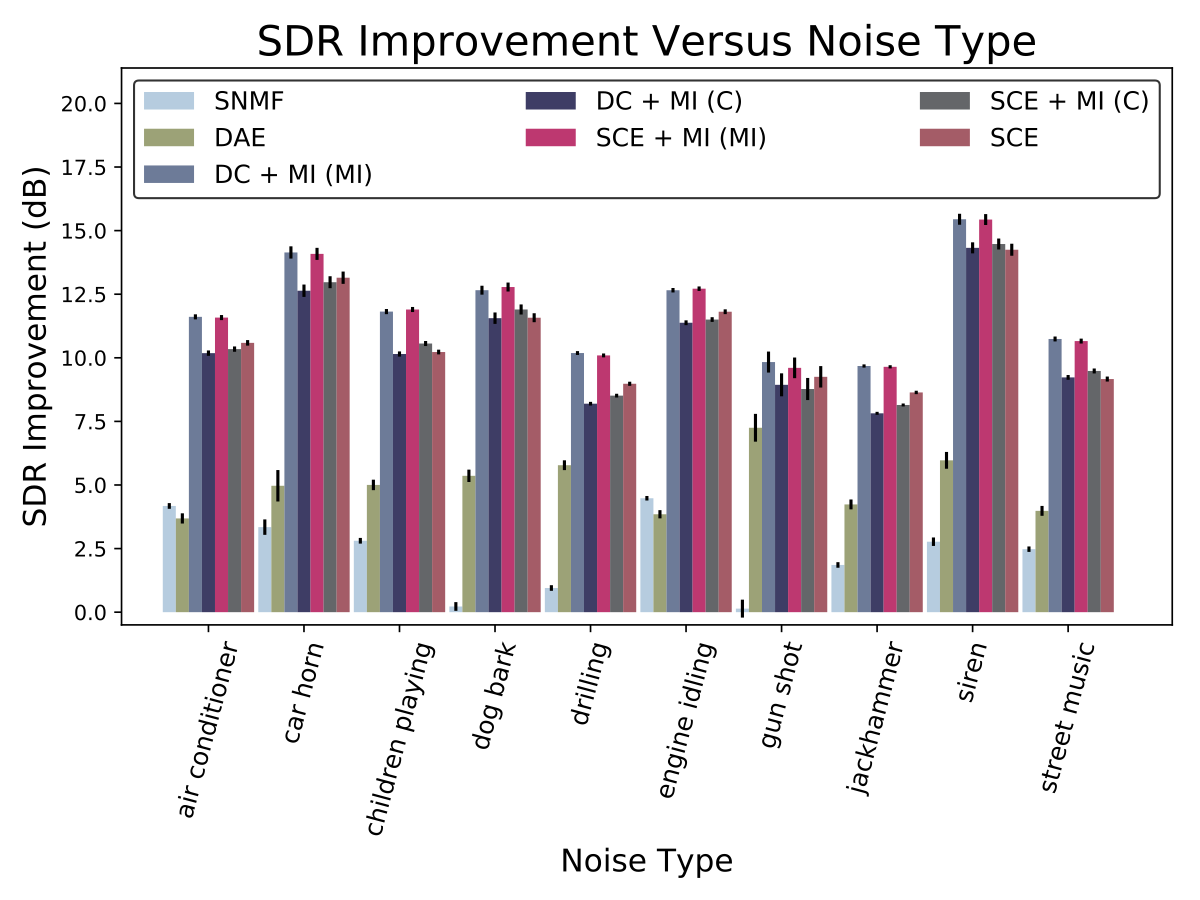}
		\caption{Performance versus noise source. DC+MI (C) represents using the embeddings for clustering to reconstruct a binary mask. DC+MI (MI) represents using mask inference for source separation. Likewise for SCE+MI (C) and (MI).} 
		\label{OutOfSetSDRDeltaVersusNoiseType}
	\end{subfigure}
	\vspace{1em} 
	\begin{subfigure}{0.4\textwidth} 
		\includegraphics[width=\textwidth]{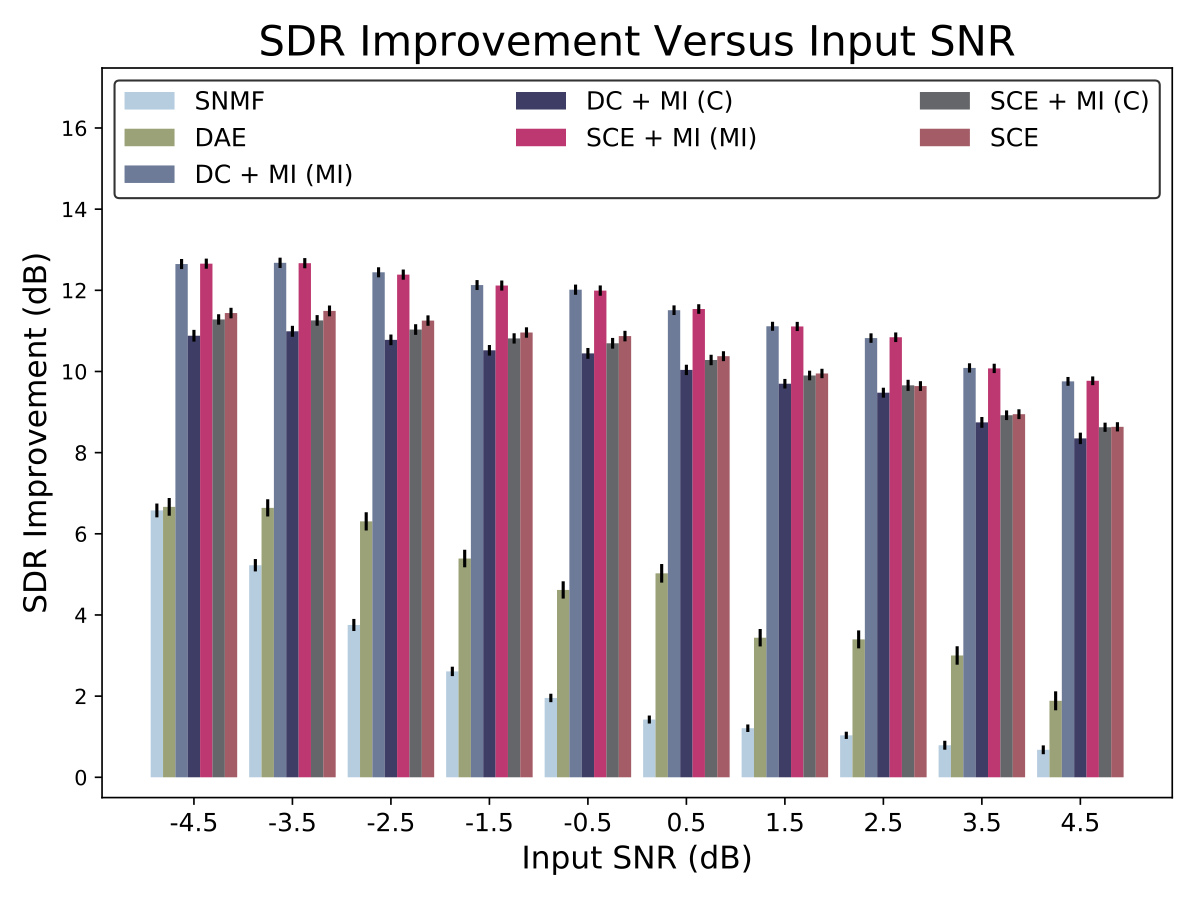}
		\caption{Performance versus input signal-to-noise ratio (SNR). The values indicated for the input SNR represent a range of SNRs $\pm.5\mathrm{dB}$ around the shown value. (i.e. $3.5\mathrm{dB}$ represents SNRs in the range of $\left[3,4\right]\mathrm{dB}$)}
		\label{OutOfSetSDRDeltaVersusInputSNR}
	\end{subfigure}
\end{figure}

The results from our experiments on a hold-out set of mixes are summarized in Figs.~\ref{OutOfSetSDRDeltaVersusNoiseType} and~\ref{OutOfSetSDRDeltaVersusInputSNR}.
The performance of the mask inference head of SCE+MI is on par with the DC+MI (+13 dB at an input SNR of [-5,-4] dB) while the clustering performance of SCE (+11.5 dB) is slightly better than the clustering of the SCE+MI and DC+MI algorithms.
Thus, SCE may be more desirable when the number of sources to be separated is arbitrary.
The improvements in SDR are greatest for more statistically stationary noise sources and inputs with lower in SNR.
This can be explained by the fact that at higher input SNRs the signal is already quite prominent, so there is less room for improvement.
The performance of the deep learning-based methods is relatively consistent across input SNRs while SNMF sees more dramatic differences.
%
%
\section{Conclusions}
We show that SCE with mask inference gives improved reconstruction performance for dynamic noise source denoising. Mask inference performs well (on average, +12 dB in SDR) regardless of the clustering loss it's coupled with. SCE showed the best clustering performance (on average, +11 dB in SDR). This indicates that denoising in the presence of an arbitrary number of sources, SCE may give better accuracy.

At present, the training objective related to the embedding space (SCE or DC) is not perfectly aligned with the $K$-means clustering performed at inference. Immediate future work on incorporating the $k$-means objective into the training procedure~\cite{2016arXiv161004794Y} could improve clustering performance.

\vfill\pagebreak



\bibliographystyle{interspeech-2018/IEEEtran}
\bibliography{main}

\begin{thebibliography}{10}
\providecommand{\url}[1]{#1}
\csname url@samestyle\endcsname
\providecommand{\newblock}{\relax}
\providecommand{\bibinfo}[2]{#2}
\providecommand{\BIBentrySTDinterwordspacing}{\spaceskip=0pt\relax}
\providecommand{\BIBentryALTinterwordstretchfactor}{4}
\providecommand{\BIBentryALTinterwordspacing}{\spaceskip=\fontdimen2\font plus
\BIBentryALTinterwordstretchfactor\fontdimen3\font minus
  \fontdimen4\font\relax}
\providecommand{\BIBforeignlanguage}[2]{{%
\expandafter\ifx\csname l@#1\endcsname\relax
\typeout{** WARNING: IEEEtran.bst: No hyphenation pattern has been}%
\typeout{** loaded for the language `#1'. Using the pattern for}%
\typeout{** the default language instead.}%
\else
\language=\csname l@#1\endcsname
\fi
#2}}
\providecommand{\BIBdecl}{\relax}
\BIBdecl

\bibitem{bss-beamforming}
\BIBentryALTinterwordspacing
J.~Sanz-Robinson, L.~Huang, T.~Moy, W.~Rieutort-Louis, Y.~Hu, S.~Wagner, J.~C.
  Sturm, and N.~Verma, ``Robust blind source separation in a reverberant room
  based on beamforming with a large-aperture microphone array.'' in
  \emph{ICASSP}.\hskip 1em plus 0.5em minus 0.4em\relax IEEE, 2016, pp.
  440--444. [Online]. Available:
  \url{http://dblp.uni-trier.de/db/conf/icassp/icassp2016.html}
\BIBentrySTDinterwordspacing

\bibitem{monaural-ica}
\BIBentryALTinterwordspacing
L.~K. Hansen and K.~B. Petersen, ``Monaural {ICA} of white noise mixtures is
  hard,'' in \emph{Proceedings of {ICA'}2003 Fourth Int. Symp.. on Independent
  Component Analysis and Blind Signal Separation, Nara Japan, April 4,}, 2003,
  pp. 815--820. [Online]. Available:
  \url{http://www2.imm.dtu.dk/pubdb/p.php?1650}
\BIBentrySTDinterwordspacing

\bibitem{sparse-nmf}
\BIBentryALTinterwordspacing
N.~Mohammadiha, P.~Smaragdis, and A.~Leijon, ``Supervised and unsupervised
  speech enhancement using nonnegative matrix factorization,'' \emph{CoRR},
  vol. abs/1709.05362, 2017. [Online]. Available:
  \url{http://arxiv.org/abs/1709.05362}
\BIBentrySTDinterwordspacing

\bibitem{le2015sparse}
J.~Le~Roux, F.~J. Weninger, and J.~R. Hershey, ``Sparse nmf--half-baked or well
  done?'' \emph{Mitsubishi Electric Research Labs (MERL), Cambridge, MA, USA,
  Tech. Rep., no. TR2015-023}, 2015.

\bibitem{wavenet}
\BIBentryALTinterwordspacing
A.~van~den Oord, S.~Dieleman, H.~Zen, K.~Simonyan, O.~Vinyals, A.~Graves,
  N.~Kalchbrenner, A.~W. Senior, and K.~Kavukcuoglu, ``Wavenet: {A} generative
  model for raw audio,'' \emph{CoRR}, vol. abs/1609.03499, 2016. [Online].
  Available: \url{http://arxiv.org/abs/1609.03499}
\BIBentrySTDinterwordspacing

\bibitem{hershey2016deep}
J.~R. Hershey, Z.~Chen, J.~Le~Roux, and S.~Watanabe, ``Deep clustering:
  Discriminative embeddings for segmentation and separation,'' in
  \emph{Acoustics, Speech and Signal Processing (ICASSP), 2016 IEEE
  International Conference on}.\hskip 1em plus 0.5em minus 0.4em\relax IEEE,
  2016, pp. 31--35.

\bibitem{2016chimera}
Y.~{Luo}, Z.~{Chen}, J.~R. {Hershey}, J.~{Le Roux}, and N.~{Mesgarani}, ``{Deep
  Clustering and Conventional Networks for Music Separation: Stronger
  Together},'' \emph{ArXiv e-prints}, Nov. 2016.

\bibitem{mikolov}
\BIBentryALTinterwordspacing
T.~Mikolov, I.~Sutskever, K.~Chen, G.~S. Corrado, and J.~Dean, ``Distributed
  representations of words and phrases and their compositionality,'' in
  \emph{Advances in Neural Information Processing Systems 26}, C.~J.~C. Burges,
  L.~Bottou, M.~Welling, Z.~Ghahramani, and K.~Q. Weinberger, Eds.\hskip 1em
  plus 0.5em minus 0.4em\relax Curran Associates, Inc., 2013, pp. 3111--3119.
  [Online]. Available:
  \url{http://papers.nips.cc/paper/5021-distributed-representations-of-words-and-phrases-and-their-compositionality.pdf}
\BIBentrySTDinterwordspacing

\bibitem{gutmann2010noise}
M.~Gutmann and A.~Hyv{\"a}rinen, ``Noise-contrastive estimation: A new
  estimation principle for unnormalized statistical models,'' in
  \emph{Proceedings of the Thirteenth International Conference on Artificial
  Intelligence and Statistics}, 2010, pp. 297--304.

\bibitem{Kameoka2016}
\BIBentryALTinterwordspacing
H.~Kameoka, \emph{Non-negative Matrix Factorization and Its Variants for Audio
  Signal Processing}.\hskip 1em plus 0.5em minus 0.4em\relax Tokyo: Springer
  Japan, 2016, pp. 23--50. [Online]. Available:
  \url{https://doi.org/10.1007/}
\BIBentrySTDinterwordspacing

\bibitem{monaural-nmf}
T.~Virtanen, ``Monaural sound source separation by nonnegative matrix
  factorization with temporal continuity and sparseness criteria,'' \emph{IEEE
  Transactions on Audio, Speech, and Language Processing}, vol.~15, no.~3, pp.
  1066--1074, March 2007.

\bibitem{schmidt-nmf}
M.~N. Schmidt, ``Speech separation using non-negative features and sparse
  non-negative matrix factorization,'' in \emph{Computer Speech and Language,
  2008, submitted. [Online]. Available:
  http://www.imm.dtu.dk/pubdb/p.php?5377}, 2008.

\bibitem{feng2014speech}
X.~Feng, Y.~Zhang, and J.~Glass, ``Speech feature denoising and dereverberation
  via deep autoencoders for noisy reverberant speech recognition,'' in
  \emph{Acoustics, Speech and Signal Processing (ICASSP), 2014 IEEE
  International Conference on}.\hskip 1em plus 0.5em minus 0.4em\relax IEEE,
  2014, pp. 1759--1763.

\bibitem{2017arXiv170308019G}
E.~M. {Grais} and M.~D. {Plumbley}, ``{Single Channel Audio Source Separation
  using Convolutional Denoising Autoencoders},'' \emph{ArXiv e-prints}, Mar.
  2017.

\bibitem{vincent2010stacked}
P.~Vincent, H.~Larochelle, I.~Lajoie, Y.~Bengio, and P.-A. Manzagol, ``Stacked
  denoising autoencoders: Learning useful representations in a deep network
  with a local denoising criterion,'' \emph{Journal of Machine Learning
  Research}, vol.~11, no. Dec, pp. 3371--3408, 2010.

\bibitem{isik16}
Y.~Isik, J.~L. Roux, Z.~Chen, S.~Watanabe, and J.~R. Hershey, ``Single-channel
  multi-speaker separation using deep clustering,'' \emph{CoRR,
  http://arxiv.org/abs/1607.02173}, vol. abs/1607.02173, 2016.

\bibitem{attractor}
\BIBentryALTinterwordspacing
Z.~Chen, Y.~Luo, and N.~Mesgarani, ``Deep attractor network for
  single-microphone speaker separation,'' \emph{CoRR}, vol. abs/1611.08930,
  2016. [Online]. Available: \url{http://arxiv.org/abs/1611.08930}
\BIBentrySTDinterwordspacing

\bibitem{dong-permutation}
\BIBentryALTinterwordspacing
D.~Yu, M.~Kolb{\ae}k, Z.~Tan, and J.~Jensen, ``Permutation invariant training
  of deep models for speaker-independent multi-talker speech separation,''
  \emph{CoRR}, vol. abs/1607.00325, 2016. [Online]. Available:
  \url{http://arxiv.org/abs/1607.00325}
\BIBentrySTDinterwordspacing

\bibitem{librispeech}
V.~Panayotov, G.~Chen, D.~Povey, and S.~Khudanpur, ``Librispeech: an asr corpus
  based on public domain audio books,'' in \emph{Acoustics, Speech and Signal
  Processing (ICASSP), 2015 IEEE International Conference on}.\hskip 1em plus
  0.5em minus 0.4em\relax IEEE, 2015, pp. 5206--5210.

\bibitem{UrbanSound2014}
J.~Salamon, C.~Jacoby, and J.~P. Bello, ``A dataset and taxonomy for urban
  sound research,'' in \emph{22st {ACM} International Conference on Multimedia
  ({ACM-MM'14})}, Orlando, FL, USA, Nov. 2014.

\bibitem{tensorflow}
\BIBentryALTinterwordspacing
M.~Abadi, A.~Agarwal, P.~Barham, E.~Brevdo, Z.~Chen, C.~Citro, G.~S. Corrado,
  A.~Davis, J.~Dean, M.~Devin, S.~Ghemawat, I.~Goodfellow, A.~Harp, G.~Irving,
  M.~Isard, Y.~Jia, R.~Jozefowicz, L.~Kaiser, M.~Kudlur, J.~Levenberg,
  D.~Man\'{e}, R.~Monga, S.~Moore, D.~Murray, C.~Olah, M.~Schuster, J.~Shlens,
  B.~Steiner, I.~Sutskever, K.~Talwar, P.~Tucker, V.~Vanhoucke, V.~Vasudevan,
  F.~Vi\'{e}gas, O.~Vinyals, P.~Warden, M.~Wattenberg, M.~Wicke, Y.~Yu, and
  X.~Zheng, ``{TensorFlow}: Large-scale machine learning on heterogeneous
  systems,'' 2015, software available from tensorflow.org. [Online]. Available:
  \url{http://tensorflow.org/}
\BIBentrySTDinterwordspacing

\bibitem{wang2014training}
Y.~Wang, A.~Narayanan, and D.~Wang, ``On training targets for supervised speech
  separation,'' \emph{IEEE/ACM Transactions on Audio, Speech and Language
  Processing (TASLP)}, vol.~22, no.~12, pp. 1849--1858, 2014.

\bibitem{2016arXiv160608921M}
X.-J. {Mao}, C.~{Shen}, and Y.-B. {Yang}, ``{Image Restoration Using
  Convolutional Auto-encoders with Symmetric Skip Connections},'' \emph{ArXiv
  e-prints}, Jun. 2016.

\bibitem{2016arXiv161004794Y}
B.~{Yang}, X.~{Fu}, N.~D. {Sidiropoulos}, and M.~{Hong}, ``{Towards
  K-means-friendly Spaces: Simultaneous Deep Learning and Clustering},''
  \emph{ArXiv e-prints}, Oct. 2016.

\end{thebibliography}

\end{document}